\input mtexsis
\input epsf
\paper
\singlespaced
\widenspacing
\twelvepoint
\Eurostyletrue
\thicksize=0pt
\sectionminspace=0.1\vsize

\def\nc{{N_c}}
\def\yo1{{F_\pi^2}}
\def\llra{{\relbar\joinrel\longrightarrow}}
\def\mapright#1{{\smash{\mathop{\llra}\limits_{#1}}}}

\def\sss{\scriptscriptstyle}


\referencelist
\reference{thooft} G.~'t Hooft,
\journal Nucl. Phys. B;72,461 (1974);
\journal Nucl. Phys. B;156,269 (1979)
\endreference
\reference{Bloom}
E.D.~Bloom and F.J.~Gilman,
\journal Phys. Rev. Lett.;25,1140 (1970);\hfill\break
\journal Phys. Rev. D;4,2901 (1971)
\endreference
\reference{Poggio}
E.C.~Poggio, H.R.~Quinn and S.~Weinberg,
\journal Phys. Rev. D;13,1958 (1976)
\endreference
\reference{Isgur}
For a recent discussion, see N.~Isgur, S.~Jeschonnek, W.~Melnitchouk and J.W.~Van Orden, {{\tt hep-ph/0104022}}
\endreference
\reference{Grinstein}
B.~Grinstein and R.~F.~Lebed,
\journal Phys. Rev. D;57,1366 (1998), {{\tt hep-ph/9708396}};
B.~Blok, M.~Shifman and D-X.~Zhang,
\journal Phys. Rev. D;57,2691 (1998), {{\tt hep-ph/9709333}}
\endreference
\reference{Bramon}
A.~Bramon, E.~Etim and M.~Greco,
\journal Phys. Lett. B;41,609 (1972);
M.~Greco,
\journal Nucl. Phys. B;63,398 (1973);
J.~J.~Sakurai,
\journal Phys. Lett. B;46,207 (1973);
B.~V.~Geshkenbein,
\journal Sov. J. Nucl. Phys;49,705 (1989)
\endreference
\reference{Golterman}
M.~Golterman and S.~Peris, 
\journal JHEP;0101,028 (2001), {{\tt hep-ph/0101098}}
\endreference
\reference{Veneziano} 
G.~Veneziano,
\journal Nuovo Cim. A;57,190 (1968)
\endreference
\reference{Lovelace} C.~Lovelace, 
\journal Phys. Lett. B;28,264 (1968);\hfill\break
J.~A.~Shapiro,
\journal Phys. Rev.;179,1345 (1969)
\endreference
\reference{Coleman} S.~Coleman and E.~Witten,
\journal Phys. Rev. Lett.;45,100 (1980)
\endreference
\reference{Shifman}
See, for instance, M.~Shifman, {{\tt hep-ph/0009131}}
\endreference
\reference{Shifman2}
M.~Shifman, {{\tt hep-ph/9405246}}
\endreference
\reference{sfsr}  S.~Weinberg, \journal Phys. Rev. Lett.;18,507 (1967);
 C.~Bernard, A.~Duncan, J.~LoSecco and S.~Weinberg, 
\journal Phys. Rev. D;12,792 (1975)
\endreference
\reference{derafael} E.~de Rafael and M.~Knecht, 
\journal Phys. Lett. B;424,335 (1998), {{\tt hep-ph/9712457}}
\endreference
\reference{algebraic}  S.~Weinberg, \journal Phys. Rev.;177,2604 (1969);\hfill\break
S.~Weinberg, \journal Phys. Rev. Lett.;65,1177 (1990)
\endreference
\reference{Beane1}
S.~R.~Beane,
\journal J. Phys. G;27,727 (2001), {{\tt hep-ph/9909571}}
\endreference
\reference{pipi} {{\it Pion-pion Interactions in Particle Physics}},
by B.R.~Martin, D.~Morgan and G.~Shaw, (Academic Press, London, 1976)
\endreference
\reference{Ananthanarayan}
B.~Ananthanarayan, G.~Colangelo, J.~Gasser and H.~Leutwyler, {{\tt hep-ph/0005297}}
\endreference
\reference{Ademollo}  M.~Ademollo, G.~Veneziano, and S.~Weinberg, 
                  \journal Phys. Rev. Lett.;22,83 (1969)
\endreference
\reference{Nambu} 
Y.~Nambu and P.~Frampton, in
{{\it Essays In Theoretical Physics Dedicated To G. Wentzel}}, (Chicago, 1970)
\endreference
\reference{Dominguez}
C.~A.~Dominguez, {{\tt hep-ph/0102190}}
\endreference
\reference{Srivastava}
E.~Etim, M.~Greco and Y.~Srivastava,
\journal Lett. Nuovo Cim.;16,65 (1976)
\endreference
\reference{Peris}
S.~Peris, B.~Phily and E.~de Rafael,
\journal Phys. Rev. Lett.;86,14 (2001), {{\tt hep-ph/0007338}}
\endreference
\reference{Beane2}
S.R.~Beane,
\journal Phys. Rev. D;61,116005 (2000), {{\tt hep-ph/9910525}}
\endreference
\reference{pdg}
D.E.~Groom {\it et al.}  
\journal Eur. Phys. J. C;15,1 (2000)
\endreference
\reference{Indu}
F.J.~Yndurain,
\journal Phys. Rept.;320,287 (1999), {{\tt hep-ph/9903457}}
\endreference
\reference{Stern}
M.~Davier, L.~Girlanda, A.~Hocker and J.~Stern,
\journal Phys. Rev. D;58,096014 (1998), {{\tt hep-ph/9802447}}
\endreference
\reference{Meyer}
C.A.~Meyer, {\it private communication}
\endreference
\reference{Glozman}
T.D.~Cohen and L.Ya.~Glozman, {{\tt hep-ph/0102206}}
\endreference
\endreferencelist

\titlepage
\obeylines
\hskip4.8in{NT@UW-01-012}\unobeylines
\vskip0.5in
\title
Constraining Quark-Hadron Duality at Large-$\nc$ 
\endtitle
\author
Silas R.~Beane

Department of Physics, University of Washington
Seattle, WA 98195-1560
\vskip0.1in
\center{{\it sbeane@phys.washington.edu}}\endcenter
\endauthor

\abstract
\singlespaced
\widenspacing

Quark-meson duality for two-point functions of vector and axial-vector QCD
currents is investigated in the large-$\nc$ approximation. We find that the joint
constraints of duality and chiral symmetry imply degeneracy of excited vector
and axial-vector mesons in the large-$\nc$ limit. We compare model-independent
constraints with expectations based on the Veneziano-Lovelace-Shapiro string
model. Several models of duality are constructed, and phenomenological
implications are discussed.

\endabstract 
\vskip0.5in
\endtitlepage
\vfill\eject                                     
\superrefsfalse
\singlespaced
\widenspacing

\vskip0.1in
\noindent {\twelvepoint{\bf 1.\quad Introduction}}
\vskip0.1in

\noindent In the large-$\nc$ limit, QCD correlators of quark bilinears can be
expressed as sums of zero-width meson tree graphs\ref{thooft}. These sums must
be infinite in order to be consistent with perturbative QCD logarithms at large
momentum transfer. The detailed matching of hadronic and partonic degrees of
freedom, known as quark-hadron duality\ref{Bloom}\ref{Poggio}\ref{Isgur}, has
been explicitly verified in QCD in 1+1 dimensions in the large-$\nc$
limit\ref{Grinstein}.  It has long been thought that the exchange of an
infinite number of vector mesons is in some sense dual to the perturbative QCD
continuum\ref{Bramon}. Early work uncovered various intriguing similarities
between the simplest models of quark-meson duality and hadronic string
models. Given the widespread belief that large-$\nc$ QCD is in some sense
equivalent to a string theory, these similarities have received recent
attention\ref{Golterman}. Following \Ref{Golterman}, in this paper we
investigate duality in the large-$\nc$ limit in the simplest correlators that
have an operator product expansion (OPE); i.e. two-point functions of vector
and axial-vector currents. We point out that there are non-trivial chiral
symmetry constraints which must be satisfied in addition to those constraints
implied by duality. We discuss the interesting dilemma raised by simultaneous
satisfaction of all constraints. These constraints suggest that there is an
infinite tower of {\it degenerate} vector and axial-vector mesons in the
large-$\nc$ limit. The phenomenological implications of this conjecture are
considered in a simple model.  As an example of a system with an infinite
spectrum of mesons we consider how chiral symmetry is satisfied in the
Lovelace-Shapiro-Veneziano (LSV) string model\ref{Veneziano}\ref{Lovelace} and
we investigate the implications of that model for duality.

\vskip0.1in
\noindent {\twelvepoint{\bf 2.\quad Duality Constraints}}
\vskip0.1in

\noindent In large-$\nc$ QCD, mesons have the most general quantum numbers of
the quark bilinear ${\bar {\rm q}}\Gamma {\rm q}$ where $\Gamma$ is some
arbitrary spin structure\ref{thooft}.  Hence all mesons have zero or unit
isospin and transform as $(\bf{2},\bf{2})$, $(\bf{3},\bf{1})$ and/or
$(\bf{1},\bf{3})$ with respect to $SU({2})\times SU({2})$ (to be precise,
large-$\nc$ QCD has a $U(2)\times U(2)$ chiral symmetry).  We will assume that
the order parameter of chiral symmetry breaking in QCD with two massless
flavors transforms as $(\bf{2},\bf{2})$. Assuming confinement, it then follows
that chiral symmetry is spontaneously broken\ref{Coleman}. The conserved vector
and axial-vector currents, ${{\rm V}_{\mu}^a}$ and ${{\rm A}_{\mu}^a}$ form a
six-dimensional multiplet; hence they transform as $(\bf{3},\bf{1})\oplus (\bf{1},\bf{3})$.  
Consider the time-ordered product of vector currents

$$ {\Pi_{\sss VV}^{\mu\nu}}(q){\delta_{ab}}=2i\int {d^4x}{e^{iqx}}
\bra{0}T\lbrack{{\rm V}_{a}^\mu}(x){{\rm V}_{b}^\nu}(0)\rbrack\ket{0}. 
\EQN correl1$$ 
Here ${\Pi_{\sss VV}}$ transforms as
$(\bf{1},\bf{1})\oplus(\bf{3},\bf{3})\oplus\ldots$ with respect to 
$SU(2)\times SU(2)$.  Lorentz invariance and current conservation allow the decomposition

$$ {\Pi_{\sss VV}^{\mu\nu}}(q)= ({q^\mu}{q^\nu}-{g^{\mu\nu}}{q^2}){\Pi_{\sss
V}}({Q^2}), 
\EQN correl2$$ 
where ${Q^2}=-{q^2}$. Identical considerations for the $AA$ correlator lead to
$\Pi_{\sss A}({Q^2})$.  One can write a dispersive representation of the
function $\Pi_{\sss {V,A}}({Q^2})$ and saturate with an infinite number of
zero-width meson states. This dispersion relation requires one subtraction,
however we will assume an unsubtracted dispersion relation and track the
divergent part. We find, in the large-$\nc$ limit,

$$\EQNalign{ 
&{\Pi_{\sss V}}({Q^2})=
2{\sum _{n=0}^\infty}
{{{F_{\sss V}^2}(n)}\over{{Q^2}+{M_{\sss V}^2}(n)}}\; ,
\EQN vectorsum;a\cr
&{\Pi_{\sss A}}({Q^2})=2{{F_{\pi}^2}\over{Q^2}}+
2{\sum _{n=0}^\infty}
{{{F_{\sss A}^2}(n)}\over{{Q^2}+{M_{\sss A}^2}(n)}},
\EQN vectorsum;b\cr}
$$
where ${F_{\sss {V,A}}}(n)$ and ${M_{\sss {V,A}}}(n)$ are the vector and
axial-vector decay constants and masses, respectively.
Because the functions ${\Pi_{\sss {V,A}}}({Q^2})$ transform as 
$(\bf{1},\bf{1})$,
they have perturbative components which are easily computed in QCD
perturbation theory. The Euler-Maclaurin summation formula
implies the duality matching condition

$$
2{\int _{0}^\infty}{dn}
{{{F_{\sss {V,A}}^2}(n)}\over{{Q^2}+{M_{\sss {V,A}}^2}(n)}}+{\cal O}(Q^{-2})
\, \, \mapright{{Q^2}\rightarrow\infty}\, \,
-{\nc\over{12{\pi^2}}}\;\log{Q^2}+\ldots +{\sum_{m=1}^{\infty}} 
{{\vev{{\cal O}}^{\sss d=2m}_{\sss {V,A}}}\over{Q^{2m}}}
\EQN euler-mac$$ 
where the dots correspond to the logarithmic divergence which appears on both sides
of the equation and the $\vev{{\cal O}}$'s are Wilson coefficients of mass-dimension $d$.
The coefficient of the logarithm is computed in perturbative QCD\ref{Shifman}\ref{Golterman}.
The duality matching condition implies 

$$
{{{F_{\sss {V,A}}^2}(n)}/{{M_{\sss {V,A}}^2}(n)}}\quad\mapright{{n\,\rightarrow\,\infty}}
\quad{n^{-1}}.
\EQN limt$$ 
In addition to this asymptotic constraint, there are constraints on the
$n$ dependences of the couplings and masses: (i) the existence of an OPE implies
that the sums over $n$ in \Eq{vectorsum} must generate functions which, aside
from perturbative logarithms, are analytic in $1/Q^{2}$; (ii) the coefficients
of the OPE must have factorial behavior in $n$$^1$\vfootnote1{\tenpoint
\Ref{Shifman2} points out that for ${{F_{\sss {V,A}}^2}(n)}={F_{\sss {V,A}}^2}$,
the sums over $n$ in \Eq{vectorsum} are Euler $\psi$-functions which satisfy
(i) and (ii). The occurence of gamma functions is reminiscent of
hadronic string models.}; (iii) chiral symmetry must be preserved. We will
address the issue of chiral symmetry in detail in the next section.

\vfill\eject
\vskip0.1in
\noindent {\twelvepoint{\bf 3.\quad Chiral Constraints}}
\vskip0.1in

\vskip0.1in
\noindent {\twelvepoint{\it 3.1\quad Matching to Perturbation Theory}}

\noindent In the ${Q^2}\rightarrow\infty$ limit, duality dictates that the
infinite sums over vector and axial-vector meson states
match to a perturbative expansion in $\alpha_s$. This expansion is defined
in the asymptotically-free phase where chiral symmetry is unbroken. Therefore,
in the matching region, each vector meson in the infinite sum must be paired
with a {\it degenerate} axial-vector chiral partner;
pair-by-pair they fill out {\it irreducible}
$(\bf{1},\bf{3})\oplus(\bf{3},\bf{1})$ representations of the chiral group.
This leads to the asymptotic constraints

\offparens
$$\EQNalign{ 
&{{{F_{\sss {V}}^2}(n)}/{{F_{\sss {A}}^2}(n)}}\quad\mapright{{n\,\rightarrow\,\infty}}\quad{1}\; ,
\EQN pt;a\cr
&{{{M_{\sss {V}}^2}(n)}/{{M_{\sss {A}}^2}(n)}}\quad\mapright{{n\,\rightarrow\,\infty}}\quad{1}.
\EQN pt;b\cr}
$$ \autoparens 
We will see that these constraints are naturally incorporated in more general
statements of chiral symmetry which will be derived below. Notice that if 
$M_{\sss {V,A}}^2(n)$ is linear in $n$, \Eq{pt;b} implies a ``universal''
slope parameter.
\vskip0.1in
\noindent {\twelvepoint{\it 3.2\quad Matching to the OPE}}

\noindent The procedures of expanding in $1/Q^{2}$ and summing over $n$ in
${\Pi_{\sss{V,A}}}({Q^2})$ do not commute. This is due to the presence of
logarithms which reorder the $1/Q^{2}$ expansion. Matching to the OPE must be
achieved by summing over $n$ and only then expanding in $1/Q^{2}$. However,
this non-commutativity is {\it not true} of the correlator

$$
{\Pi_{\sss {LR}}}({Q^2})\equiv {1\over 2}({\Pi_{\sss {V}}}({Q^2})-{\Pi_{\sss {A}}}({Q^2}))
\, \, \mapright{{Q^2}\rightarrow\infty}\, \,
{\sum_{m=1}^{\infty}} 
{{\vev{{\cal O}}^{\sss d=2m}_{\sss \bf{(3,3)}}}\over{Q^{2m}}}.
\EQN correl3$$ 
The subscript labeling the Wilson coefficients indicates that this correlator
transforms as $(\bf{3},\bf{3})$ and therefore contains no perturbative
logarithm. Hence performing the sum over $n$ does not rearrange the $1/{Q^2}$
expansion, and one can expand in $1/{Q^2}$ {\it before} performing the infinite
sum over $n$. We will see in the next section that this commutativity is
protected by chiral symmetry. Since the first two OPE coefficients in \Eq{correl3} 
vanish in QCD in the chiral limit, one reads directly
from \Eq{vectorsum} the spectral-function sum rules in the large-$\nc$ limit\ref{sfsr}\ref{derafael}

\offparens
$$\EQNalign{ 
&{\sum _{n=0}^\infty}{F_{\sss V}^2}(n)-{\sum _{n=0}^\infty}{F_{\sss A}^2}(n)=\yo1\; ,
\EQN sfsr;a\cr
&{\sum _{n=0}^\infty}{F_{\sss V}^2}(n){M_{\sss V}^2}(n)
-{\sum _{n=0}^\infty}{F_{\sss A}^2}(n){M_{\sss A}^2}(n)=0.
\EQN sfsr;b\cr}
$$ \autoparens 
These sum rules must be satisfied by any model of large-$\nc$ QCD consistent
with chiral symmetry. The asymptotic constraints of \Eq{pt} are enforced by
these sum rules.

\vskip0.1in
\noindent {\twelvepoint{\it 3.3\quad Constraints from the Chiral Algebra}}

\noindent It will prove useful to give a derivation of \Eq{sfsr} which is independent of
the OPE\ref{algebraic}\ref{Beane1} as it will allow contact with hadronic
string models. For this purpose, it is convenient to work in the infinite
momentum frame. This is a natural choice given our interest in large
(Euclidean) momenta. Of course, the results that we derive are true in all
frames. A useful property of the infinite-momentum frame is that the axial 
charges annihilate the vacuum, ${Q_{5}^{a}}\ket{0}=0$. If we boost the vector 
mesons along the $3$-axis to ${p_\mu}=({p_0},0,0,{p_3})$, we can write,
in the ${p_3}\rightarrow\infty$ limit,

\offparens
$$\EQNalign{ &\bra{0}{{\rm A}_{\mu}^a}\ket{\pi^b}=
{\delta^{ab}}{F_{\pi}}{p_\mu}  \; ,\EQN decays;a\cr 
&\bra{0}{{\rm A}_{\mu}^a}\ket{A^{b}}^{\sss (0)}= 
{\delta^{ab}}{F_{\sss A}}{M_{\sss A}}{\epsilon_\mu^{\sss (0)}}
={\delta^{ab}}{F_{\sss A}}{p_\mu}+{\cal O}({{p^{-1}_3}})\; ,\EQN decays;b\cr 
&\bra{0}{{\rm V}_{\mu}^a}\ket{V^{b}}^{\sss (0)}= 
{\delta^{ab}}{F_{\sss V}}{M_{\sss V}}{\epsilon_\mu^{\sss (0)}}
={\delta^{ab}}{F_{\sss V}}{p_\mu}+{\cal O}({{p^{-1}_3}}),\EQN decays;c\cr} 
$$ 
where the superscripts in parentheses label the helicity, $\lambda$.
It will prove worthwhile to consider matrix elements of the axial charges
as well. The matrix element for a transition from a meson $\beta$ to a meson
$\alpha$ and a pion in the infinite-momentum frame is given by

\autoparens
$$ {{\cal M}}(\beta({\lambda '})\rightarrow \alpha({\lambda })+\pi_a)
={{({{F_\pi}})}^{-1}} ({M_\alpha ^2}-{M_\beta ^2})
{^{\sss (\lambda ')}\bra{\,\beta\,}}
{Q^{\sss 5}_a}{{\ket{\,\alpha\,}}^{\sss (\lambda )}}
{\delta _{{\lambda '}\lambda}},
\EQN amp $$\autoparens
where the Kronecker delta ensures helicity conservation and
${Q^{\sss 5}_a}$ is a conserved axial charge. We define

$$\EQNalign{ 
\bra{\pi_b}{Q^{\sss 5}_{a}}{\ket{S}}&=
-i{\delta_{ab}}{G_{{\sss {S}}{\pi}}}/{F_\pi}\; ,
\EQN coupgenrel;a\cr
\bra{\pi_b}{Q^{\sss 5}_{a}}{\ket{V_c}}&=
-i{\epsilon_{abc}}{G_{{\sss {V}}{\pi}}}/{F_\pi}\; ,
\EQN coupgenrel;b\cr
\bra{A_b}{Q^{\sss 5}_{a}}{\ket{V_c}}&=
-i{\epsilon_{abc}}{G_{{\sss {V}}{\sss A}}}/{F_\pi}.
\EQN coupgenrel;c\cr}
$$
Here $S$, $V$ and $A$ represent meson states with 
$I^G(J^{PC})$ given by 
$0^+({\rm even}^{++})$,
$1^+({\rm odd}^{--})$ and
$1^-({\rm odd}^{++})$, respectively. We suppress the helicity labels
on the states as we are interested only in zero-helicity transitions.

Consider the following matrix elements of the chiral algebra

\offparens
$$\EQNalign{
&\bra{0}[{Q_{\sss 5}^{a}},{{\rm V}_{\mu}^b}]{\ket{A^e}}
=i{\epsilon^{abc}}\bra{0}{{\rm A}_{\mu}^c}{\ket{A^e}}\; ,
\EQN currentalg1A;a\cr
&\bra{0}[{Q_{\sss 5}^{a}},{{\rm A}_{\mu}^b}]{\ket{V^e}}
=i{\epsilon^{abc}}\bra{0}{{\rm V}_{\mu}^c}{\ket{V^e}}\; ,
\EQN currentalg1A;b\cr
&\bra{0}[{Q_{\sss 5}^{a}},{{\rm V}_{\mu}^b}]{\ket{\pi^e}}
=i{\epsilon^{abc}}
\bra{0}{{\rm A}_{\mu}^c}{\ket{\pi^e}}\; ,
\EQN currentalg1A;c\cr
&{\bra{\pi_e}}
[{Q^{\sss 5}_{a}},{Q^{\sss 5}_{b}}]{\ket{\pi_d}
=i{\epsilon_{abc}}{\bra{\pi_e}}}{T_c}
{\ket{\pi_d}}.
\EQN currentalg1A;d\cr}
$$\autoparens 
By inserting a complete set of states in the commutators and
using \Eq{decays}, \Eq{coupgenrel} and
${\bra{\pi_a}}{T_b}{\ket{\pi_c}}=i\epsilon_{abc}$,
it is easy to derive
a cornucopia of sum rules\ref{Beane1}. Consider, as an example, 
\Eq{currentalg1A;a}; there is a sum rule for each axial-vector state, labeled
by $n'$. Using ${Q_{5}^{a}}\ket{0}=0$ and inserting a complete set of states yields

\offparens
$$
-{\sum _{n=0}^\infty}\bra{0} {{\rm
V}_{\mu}^b}\ket{V^f;n}\delta_{J,1}\bra{V^f;n} 
{Q_{\sss 5}^{a}}{\ket{A^e;n'}}
=i{\epsilon^{abc}}\bra{0}{{\rm A}_{\mu}^c}{\ket{A^e;n'}},
\EQN ex1
$$\autoparens 
where the Kronecker delta constrains the sum to spin-one $V$ states.
Using \Eq{decays} and \Eq{coupgenrel}, it is easy to derive

$$
{\sum_{n=0}^\infty}{F_{\sss V}}(n)
{G^{\sss J=1}_{{\sss {V}}{\sss A}}}(n,n')={F_\pi}{F_{\sss A}}(n'),
\EQN ex1
$$\autoparens 
where the superscript indicates that the sum is over spin-one $V$ states.
The sum rules from \Eq{currentalg1A} which are of relevance to this paper are

\offparens
$$\EQNalign{ 
&{\sum _{n=0}^\infty}{F_{\sss V}^2}(n)-{\sum _{n=0}^\infty}{F_{\sss A}^2}(n)=\yo1\; ,
\EQN summrelfir;a\cr
&{\sum_{n=0}^\infty}{F_{\sss V}}(n){G^{\sss J=1}_{{\sss V}\pi}}(n)=\yo1\; ,
\EQN summrelfir;b\cr
&{\sum_{n=0}^\infty}{G^2_{{\sss S}\pi}}(n)
+{\sum_{n=0}^\infty}{G^2_{{\sss V}\pi}}(n)=\yo1 .
\EQN summrelfir;c\cr}
$$ \autoparens 
The first sum rule is the first spectral-function sum rule. We now see that, in
the large-$\nc$ limit, this sum rule is true independent of the OPE; it is a
simple consequence of chiral symmetry, which is encoded in the commutators of
\Eq{currentalg1A}. The second and third sum rules
constrain the pion vector form factor and $\pi-\pi$ scattering,
respectively\ref{Beane1}.

There are additional sum rules which involve the meson masses, and which can be
derived without the OPE\ref{Beane1}. These sum rules require the assumption
that the order parameter of chiral symmetry breaking transforms purely as
$(\bf{2},\bf{2})$. Those of relevance here are

\offparens
$$\EQNalign{ 
&{\sum _{n=0}^\infty}{F_{\sss V}^2}(n){M_{\sss V}^2}(n)
-{\sum _{n=0}^\infty}{F_{\sss A}^2}(n){M_{\sss A}^2}(n)=0\; ,
\EQN massrules;a\cr
&{\sum_{n=0}^\infty}{G^2_{{\sss S}\pi}}(n){M_{\sss S}^2}(n)
-{\sum_{n=0}^\infty}{G^2_{{\sss V}\pi}}(n){M_{\sss V}^2}(n)=0.
\EQN massrules;b\cr}
$$ \autoparens 
The first sum rule is the second spectral-function sum rule. The second sum
rule constrains $\pi-\pi$ scattering\ref{Beane1}.

\vskip0.1in
\noindent {\twelvepoint{\bf 4.\quad Trouble with Mass Splittings}}
\vskip0.1in

\noindent In this section, we consider how duality and chiral symmetry constrain
the meson decay constants and masses when 
${M_{\sss {V,A}}^2}(n)$ is a
linear function of $n$. The duality constraint, \Eq{limt}, allows the
general parametrization

$$\EQNalign{ 
&{{F_{\sss {V}}^2}}(n)= {{F_{\sss {V}}^2}}+
\;\tilde\chi_{\sss {V}} (n)+\chi_{\sss {V}} (n)\; ,
\EQN modtheorem1;a\cr
&{{F_{\sss {A}}^2}}(n)= {{F_{\sss {A}}^2}}+
\;\tilde\chi_{\sss {A}} (n)+\chi_{\sss {A}} (n)
\EQN modtheorem1;b\cr}
$$
where $\tilde\chi_{\sss {V,A}} (n)$ and $\chi _{\sss {V,A}} (n)$ are
functions which vanish as $n\rightarrow\infty$.
The general decomposition is such that 
$\sum_{n=0}^\infty{\tilde\chi_{\sss {V,A}}}(n)$ is divergent (with no finite
part) while $\sum_{n=0}^\infty{\chi}_{\sss {V,A}}(n)$ is convergent.
The first spectral-function sum rule, \Eq{sfsr;a}, implies

$$\EQNalign{ 
{{F_{\sss V}}}&={{F_{\sss A}}}\equiv F\; ,
\EQN modtheorem2;a\cr
\tilde\chi_{\sss {V}} (n)&=\tilde\chi_{\sss {A}} (n)\equiv \tilde\chi (n)\; ,
\EQN modtheorem2;b\cr
\chi_{\sss {V}} (n)&-\chi_{\sss {A}} (n)\equiv \chi (n)\; ,
\EQN modtheorem2;c\cr
\sum_{n=0}^\infty \chi (n)&=\yo1 ,
\EQN modtheorem2;d\cr}
$$ 
while the second spectral-function sum rule, \Eq{sfsr;b}, requires

$$\EQNalign{ 
{M_{\sss {V,A}}^2}(n)&={M_{\sss {V,A}}^2}+{\Lambda^2}\, n\; ,
\EQN modtheorem3;a\cr
{\sum_{n=0}^N} n\,\chi (n)&= 
{{({M_{\sss {A}}^2}-{M_{\sss {V}}^2})}\over{\Lambda^2}}
({F^2}(N+1) + \sum_{n=0}^N \tilde\chi (n) +\sum_{n=0}^\infty \chi_{\sss {V}}(n)) 
-\yo1\,{{{M_{\sss A}^2}}\over{\Lambda^2}}
\EQN modtheorem3;b\cr}
$$ 
where ${M_{\sss {V,A}}^2}$ and ${\Lambda^2}$ are free parameters, and we
have imposed a cutoff, $N$, on the number of vector and axial-vector mesons;
\Eq{modtheorem3;b} should be satisfied in the limit $N\rightarrow\infty$.  Note
that \Eq{modtheorem2;a} and \Eq{modtheorem3;a} ensure compliance with \Eq{pt;a}
and \Eq{pt;b}, respectively. 

This parametrization illustrates the difficulty in satisfying the chiral
constraints and the duality constraints simultaneously. There would appear to
be no solution, $\chi (n)$, which satisfies \Eq{modtheorem2} and
\Eq{modtheorem3} with ${M_{\sss {V}}}\neq {M_{\sss {A}}}$. For instance, by
naive power counting, \Eq{modtheorem2;d} requires that $\chi (n)$ vanish faster
than ${n^{-1}}$ for large $n$. But with this asymptotic behavior, the sum in
\Eq{modtheorem3;b} cannot generate the linear divergence necessary to balance
the equation.  Therefore, given the assumption that ${M_{\sss {V,A}}^2}(n)$ is
linear in $n$, we find no solution to the duality and chiral constraints in the
large-$\nc$ limit with ${M_{\sss {V}}}\neq {M_{\sss {A}}}$.

If the vector and axial-vector mesons are degenerate,
${M_{\sss {V}}}={M_{\sss {A}}}\equiv {M}$, and \Eq{modtheorem3;b} becomes

$$
{\sum_{n=0}^\infty} n\,\chi (n)= -\yo1\,{{{M^2}}\over{\Lambda^2}}.
\EQN modtheorem3bdeg
$$
By naive power counting, \Eq{modtheorem3bdeg} and \Eq{modtheorem2;d} can be
satisfied simultaneously if $\chi (n)$ vanishes faster than $n^{-2}$. We will
return to the degenerate case below.

Group-theoretically the situation is as follows. As pointed out above, if the
vector and axial-vector mesons are degenerate, pair-by-pair they fill out 
{\it irreducible} $(\bf{1},\bf{3})\oplus(\bf{3},\bf{1})$ representations of the
chiral group, which is rather trivial.  In the absence of degeneracy, the
vector and axial-vector mesons generally fill out infinite-dimensional {\it
reducible} sums of $(\bf{1},\bf{3})$, $(\bf{3},\bf{1})$ and $(\bf{2},\bf{2})$
representations.

\vskip0.1in
\noindent {\twelvepoint{\bf 5.\quad The Lovelace-Shapiro-Veneziano String
Model}} \vskip0.1in

\noindent Ideally, one would like to
find a smooth ansatz for ${{F_{\sss {V,A}}^2}}(n)$ which generates
both chiral physics and perturbative physics. For vector and axial-vector
squared-masses linear in $n$ and degenerate, this involves finding the function
$\chi (n)$, which satisfies \Eq{modtheorem2;d} and \Eq{modtheorem3bdeg}.
Hadronic string models are an interesting place to look for
clues. Generally these models are interesting for large-$\nc$ QCD because there
are an infinite number of mesons exchanged$^2$\vfootnote2{\tenpoint Reviews of
this model are given in \Ref{pipi} and \Ref{Ananthanarayan}.}. Consider the
following representation of the $\pi-\pi$ scattering amplitude

$$
A(s,t,u)=-{1\over 2}\lambda\lbrace \Phi (\alpha_s,\alpha_t )+\Phi (\alpha_s,\alpha_u )
-\Phi (\alpha_t,\alpha_u )\rbrace
\EQN veneuno
$$
where

$$
\Phi (a,b)\equiv {{\Gamma (1-a ) \Gamma (1-b )}\over{\Gamma (1-a -b)}}
=(1-a-b )B(1-a, 1-b )
\EQN venedos
$$
and the linear Regge trajectory is

$$
\alpha_s=\alpha_0 +\alpha s .
\EQN venetres
$$
The parameter $\lambda$, the intercept $\alpha_0$ and the slope $\alpha$
determine scattering.  Chiral symmetry requires that the amplitude have an
Adler zero at the point $s=t=u=0$. This determines
$\alpha_0 =1/2$.  Scattering is then consistent with the low-energy theorems of
chiral symmetry if one takes $\pi\lambda\alpha =F_\pi^{-2}$. Normalizing the
Regge slope to the lightest exchanged state gives $(2\alpha)^{-1}=M_\rho^2$.
Using

$$
{\rm Im}\ \Phi (\alpha_s,\alpha_t )=-\pi \sum_{n=1}^\infty
{{\Gamma (\alpha_t +n)}\over {\Gamma (n) \Gamma (\alpha_t )}}\delta(\alpha_s -n)
\EQN venequatro
$$
it is straightforward to extract the generalized couplings and masses as a function
of $n$. We find

$$
{G^2_{{\sss V}\pi}}(n)={G^2_{{\sss S}\pi}}(n)={1\over 2}\, {\chi_{\sss LSV}}(n)
\qquad n=0,1,\ldots
\EQN venecinque
$$
where

$$
{\chi_{\sss LSV}}(n)\equiv {\yo1\over{\pi}}   
{{\Gamma ({1\over2}+n)}\over {\Gamma ({1\over2}) \Gamma
(1+n)(n+{1\over2})}}.
\EQN chidef
$$
and

$$
{M^2_{{\sss V}}}(n)={M^2_{{\sss S}}}(n)={M_\rho^2}(1+2n)\qquad n=0,1,\ldots
\EQN venecinque2
$$
The sum rule, \Eq{summrelfir;c}, is then

$$
{\sum_{n=0}^{\infty}}{G^2_{{\sss {S}}{\pi}}}(n)+
{\sum_{n=0}^{\infty}}{G^2_{{\sss {V}}{\pi}}}(n)=
{\sum_{n=0}^{\infty}}
\;{\chi_{\sss LSV}}(n)
=\yo1
\EQN veneseis
$$ 
which is indeed satisfied. The states that participate in the string amplitude
are therefore in an infinite-dimensional representation of the chiral
group. This representation includes states of all spins. Notice that the mass
sum rule, \Eq{massrules;b}, is trivially satisfied by \Eq{venecinque} and
\Eq{venecinque2}, a consequence of the fact that the amplitude with $I=2$ in
the t-channel vanishes, by construction, in the LSV 
model\ref{pipi}\ref{Ananthanarayan}.

\vskip0.1in
\noindent {\twelvepoint{\bf 6.\quad Stringy Implications for Duality}}
\vskip0.1in

\vskip0.1in
\noindent {\twelvepoint{\it 6.1\quad The LSV Model}}

\noindent The LSV model is notable in that it satisfies the chiral constraints
with an infinite number of mesons and is therefore consistent with
large-$\nc$ QCD. Given the symmetric appearance of the 
chiral sum rules in \Eq{summrelfir} one might consider ${\chi_{\sss LSV}}(n)$
as an ansatz for duality in \Eq{modtheorem1} when the vector and axial-vector mesons are 
degenerate$^3$\vfootnote3{\tenpoint
A generalization of the LSV model to pion scattering
on an arbitrary hadronic target suggests 
${M^2_{{\sss A}}}(n)-{M^2_{{\sss V}}}(n)=(2\alpha)^{-1}$\ref{Ademollo}.}. 
However, for $n$ large, ${\chi_{\sss LSV}}(n)\rightarrow n^{-3/2}$, and
therefore the sum in \Eq{modtheorem3bdeg} does not converge.
There is a further related problem with this ansatz.
The sum over $n$ is easy to do in the correlators
of \Eq{vectorsum}. For large $Q$ the resulting
functions contain fractional powers of $1/Q^{2}$ and therefore do not match to the OPE.
This is no surprise since ${\chi_{\sss LSV}}(n)$ generates Regge asymptotic
behavior in $\pi-\pi$ scattering.

The chiral sum rule, \Eq{summrelfir;b}, relates ${F_{\sss V}}(n)$ to $\pi-\pi$
scattering and thus links duality and the LSV model. Consider the ansatz

$$
{F_{\sss V}}(n){G^{\sss J=1}_{{\sss V}\pi}}(n)={\chi_{\sss LSV}}(n)\qquad n=0,1,\ldots
\EQN siansatz
$$
which satisfies \Eq{summrelfir;b}. Using the duality matching condition, \Eq{limt},
this implies $({G^{\sss J=1}_{{\sss V}\pi}}(n))^2 \rightarrow n^{-3}$ for $n$ large.
We can immediately put this to the test in the LSV model; partial-wave projection yields

$$
({G^{\sss J=1}_{{\sss V}\pi}}(n))^2={3\yo1\over{\pi}}   
(n+{1\over2})^{-4} \int_{-n}^{1\over2} 
{{\Gamma(x+n+1)}\over{\Gamma(n+1)\Gamma(x)}}
(2x-{1\over2}+n).
\EQN r1n
$$
We have not succeeded in evaluating this integral to a simple expression.
Asymptotically, one finds\ref{Nambu}

$$
({G^{\sss J=1}_{{\sss V}\pi}}(n))^2
\, \, \mapright{{n}\rightarrow\infty}\, \,
n^{-5/2}(\log{n})^{-1}
\EQN r1n2
$$
which is not (quite) consistent with \Eq{siansatz}. 

\vskip0.1in
\noindent {\twelvepoint{\it 6.2\quad A Generalization of the LSV Model}}

\noindent The success of the LSV model in incorporating the chiral symmetry constraints
suggests that it might be profitable to search for simple generalizations of 
${\chi_{\sss LSV}}(n)$ that are consistent with duality as well. 
Consider, for instance\ref{Dominguez},

$$
{\chi (n,{N_{\sss{M}}},{\alpha_0})}
\equiv \yo1\,
{{\Gamma ({N_{\sss{M}}}+{\alpha_0}) (-1)^n}
\over {\Gamma ({\alpha_0}) 
\Gamma ({N_{\sss{M}}}-n)\Gamma (1+n)(n+{\alpha_0})}}\qquad {N_{\sss{M}}}>0.
\EQN stupidansatz2
$$
For integral values of ${N_{\sss{M}}}$, 
${\chi (n,{N_{\sss{M}}},{\alpha_0})}$ vanishes for $n > {N_{\sss{M}}}$, while 
for non-integral values 
${\chi (n,{N_{\sss{M}}},{\alpha_0})}$ is non-vanishing for all $n$. Using this function
one can define a one-parameter coupling which interpolates between
a finite and an infinite number of mesons$^4$\vfootnote4{\tenpoint
\Ref{Dominguez} considers 
${F_{\sss V}}(n){G^{\sss J=1}_{{\sss V}\pi}}(n)={\chi (n,{N_{\sss{M}}},1/2)}$ as an
ansatz for the pion vector form factor. For integer values,
$N_{\sss{M}}$ counts the number of vector mesons which contribute to the form factor.
Evidently, a fit to data gives $N_{\sss{M}}\sim 1.3$ which implies an infinite
number of vector mesons.}.
Note that ${\chi (n,{1/2},{1/2})}={\chi_{\sss LSV}}(n)$. 
We now have the asymptotic behavior 

$$
{\chi(n,{N_{\sss{M}}},{\alpha_0})}\;\mapright{{n}\rightarrow\infty}\;{{n^{-{(N_{\sss{M}}+1)}}}}.
\EQN asymwhatever
$$  
Therefore ${\chi (n,{N_{\sss{M}}},{\alpha_0})}$ with
$N_{\sss{M}}>1$ serves as an ansatz for duality when the vector and axial-vector mesons are 
degenerate. In effect, we find 

$$\EQNalign{ 
\sum_{n=0}^\infty {\chi (n,{N_{\sss{M}}},{\alpha_0})}&=\yo1 
\EQN theoremstup;a\cr
\sum_{n=0}^\infty n\,{\chi (n,{N_{\sss{M}}},{\alpha_0})}&=-{{\alpha_0}}\yo1\qquad N_{\sss{M}}>1
\EQN theoremstup;b\cr}
$$ 
which is in agreement with \Eq{modtheorem2;d} and 
\Eq{modtheorem3bdeg} when ${\alpha_0}={M^2}/{\Lambda^2}$. 
With ${\alpha_0}=1/2$ one finds the spectrum, \Eq{venecinque2}, of the LSV model.
This is not really surprising since the sum rules of \Eq{theoremstup} are a 
statement of chiral symmetry, 
and the Regge intercept in \Eq{venecinque2} was fixed using chiral symmetry. 
\vskip0.1in
\noindent {\twelvepoint{\bf 7.\quad Models of Duality}}
\vskip0.1in

\vskip0.1in
\noindent {\twelvepoint{\it 7.1\quad A String-Inspired Model}}

\noindent In this section we build a model of duality which is consistent with chiral
symmetry and which has no discontinuity in $n$. The vector and
axial-vector mesons are degenerate so it has little to do with the real world.
In our model we choose $\tilde\chi (n)=0$ in
\Eq{modtheorem1}$^5$\vfootnote5{\tenpoint If, for instance,
$\tilde\chi (n)\rightarrow {n^{-1}}$ for $n$ large, its
effect on duality is to generate logarithmic corrections to the OPE
coefficients.}. An ansatz consistent with duality and chiral symmetry is

$$\EQNalign{ 
{M_{\sss {V,A}}^2}(n)&={M^2}+{\Lambda^2}\, n\; ,
\EQN modtheoremmod;a\cr
{{F_{\sss {V}}^2}}(n)&= {{F^2}}+\eta\; {\chi (n,{N_{\sss{M}}},\,{{M^2}/{\Lambda^2}})}\; ,
\EQN modtheoremmod;b\cr
{{F_{\sss {A}}^2}}(n)&= {{F^2}}+(\eta -1)\; {\chi (n,{N_{\sss{M}}},\,{{M^2}/{\Lambda^2}})}
\EQN modtheoremmod;c\cr}
$$
where $\eta$ is a free parameter and 
${N_{\sss{M}}}>1$. Inserting this ansatz into \Eq{vectorsum} and
doing the sums over $n$ yields

\autoparens
$$\EQNalign{ 
{\Pi_{\sss {V,A}}}({Q^2})&=
-{{2 F^2}\over{\Lambda^2}}\;\psi ({{{M^2}+{Q^2}}\over{\Lambda^2}}) \;+\;\ldots \cr
&+{{2\eta\yo1}\over{Q^2}}\Bigg\lbrack 1-
{\epsilon_{\sss {V,A}}}\;
{{\Gamma ({{M^2}\over{\Lambda^2}}+{N_{\sss{M}}})\Gamma 
({{{M^2}+{Q^2}}\over{\Lambda^2}})}\over
{\Gamma ({{M^2}\over{\Lambda^2}})\Gamma ({N_{\sss{M}}}+{{{M^2}+{Q^2}}\over{\Lambda^2}})}}
\Bigg\rbrack \; ,
\EQN vectorsummodel;a\cr
{\Pi_{\sss{LR}}}({Q^2})&=
-{{\yo1}\over{Q^2}}\;
{{\Gamma ({{M^2}\over{\Lambda^2}}+{N_{\sss{M}}})\Gamma 
({{{M^2}+{Q^2}}\over{\Lambda^2}})}\over
{\Gamma ({{M^2}\over{\Lambda^2}})\Gamma ({N_{\sss{M}}}+{{{M^2}+{Q^2}}\over{\Lambda^2}})}}
\EQN vectorsummodel;b\cr}
$$
where the dots represent a logarithmic divergence;
${\epsilon_{\sss {V}}}=1$ and ${\epsilon_{\sss {A}}}={(\eta -1 )/\eta}$.
At large $Q^2$ we then have

\autoparens
$$\EQNalign{ 
&{\Pi_{\sss {V,A}}}({Q^2})=
-{2{F^2}\over{{\Lambda^2}}}\log{Q^2} +\ldots
+\Bigg\lbrack{{2\eta\yo1}}-{{{2{F^2}\over{\Lambda^2}}({M^2}-{1\over 2}{\Lambda^2})}}\Bigg\rbrack 
{1\over{Q^2}}\cr
&+{{{{F^2}\over{\Lambda^2}}({M^4}-{M^2}{\Lambda^2}+{1\over 6}{\Lambda^4})}}{1\over{Q^4}}
-{2{F^2}\over{3{\Lambda^2}}}  
({M^2}-{1\over 2}{\Lambda^2})({M^2}-{\Lambda^2}){M^2}{1\over{Q^6}}\cr
&-{2\eta\;{{\epsilon_{\sss {V,A}}}\;}\yo1}
{{\Gamma ({{M^2}\over{\Lambda^2}}+{N_{\sss{M}}})}\over{\Gamma ({{M^2}\over{\Lambda^2}})}}
{{\Lambda^{{2N_{\sss{M}}}}}\over{Q^{2{N_{\sss{M}}}+2}}}
+{\cal O}(Q^{-2{N_{\sss{M}}}-4},Q^{-8})\; ,
\EQN vectorsummodelasym;a\cr
&{\Pi_{\sss LR}}({Q^2})=
{\yo1}
{{\Gamma ({{M^2}\over{\Lambda^2}}+{N_{\sss{M}}})}\over{\Gamma ({{M^2}\over{\Lambda^2}})}}
{{\Lambda^{{2N_{\sss{M}}}}}\over{Q^{2{N_{\sss{M}}}+2}}}
+{\cal O}(Q^{-2{N_{\sss{M}}}-4}).
\EQN vectorsummodelasym;b\cr}
$$
Here we see that ${N_{\sss{M}}}$ must be an integer in order to match to the OPE. 
Hence ${N_{\sss{M}}}$ counts the number of vector and axial-vector mesons which
contribute to the ${\Pi_{\sss{LR}}}({Q^2})$ correlator. In principle, one would
expect ${N_{\sss{M}}}$ to be infinite.
Taking ${N_{\sss{M}}}$ (arbitrarily) large and matching to the OPE gives

$$\EQNalign{ 
&{{F^2}}={\nc\over{24{\pi^2}}}{\Lambda^2}\; ,
\EQN opejunkmod;a\cr
&{\vev{{\cal O}}^{\sss d=2}_{\sss {V,A}}}=0={2\eta\yo1}
-{\nc\over{12{\pi^2}}}({M^2}-{1\over 2}{\Lambda^2})\; ,
\EQN opejunkmod;b\cr
&{\vev{{\cal O}}^{\sss d=4}_{\sss {V,A}}}={{\alpha_s}\over{12\pi}}\vev{G_{\mu\nu}G_{\mu\nu}}=
{\nc\over{24{\pi^2}}}({M^4}-{M^2}{\Lambda^2}+{1\over 6}{\Lambda^4})\; ,
\EQN opejunkmod;c\cr
&{\vev{{\cal O}}^{\sss d=6}_{\sss {V}}}=-{28\over 9}{\pi{\alpha_s}}\vev{{\bar
q} q}^2=-{\nc\over{36{\pi^2}}}
({M^2}-{1\over 2}{\Lambda^2})({M^2}-{\Lambda^2}){M^2}\; ,
\EQN opejunkmod;d\cr
&{\vev{{\cal O}}^{\sss d=6}_{\sss {A}}}={44\over 9}{\pi{\alpha_s}}\vev{{\bar q}
q}^2=-{\nc\over{36{\pi^2}}}
({M^2}-{1\over 2}{\Lambda^2})({M^2}-{\Lambda^2}){M^2}\; ,
\EQN opejunkmod;e\cr
&\quad\vdots \cr
&{\vev{{\cal O}}^{\sss d={2{N_{\sss{M}}}+2}}_{\sss {V}}}=
-{{2\eta\yo1}}{{\Gamma ({{M^2}\over{\Lambda^2}}+{N_{\sss{M}}})}
\over{\Gamma ({{M^2}\over{\Lambda^2}})}}
{\Lambda}^{2{N_{\sss{M}}}}\;+\ldots\; ,
\EQN opejunkmod;f\cr
&{\vev{{\cal O}}^{\sss d={2{N_{\sss{M}}}+2}}_{\sss {A}}}=
-{{2(\eta -1)\yo1}}{{\Gamma ({{M^2}\over{\Lambda^2}}+{N_{\sss{M}}})}
\over{\Gamma ({{M^2}\over{\Lambda^2}})}}
{\Lambda}^{2{N_{\sss{M}}}}\;+\ldots .
\EQN opejunkmod;g\cr}
$$ 
Since there is no local QCD operator with $d=2$, we can fix $\eta$ using
\Eq{opejunkmod;b}. For large ${N_{\sss{M}}}$, there is no solution
with $\vev{\bar q q}\neq 0$.  This is is not inconsistent with degenerate
vector and axial-vector mesons. While this model is clearly unrealistic, it 
provides an existence proof of a smooth chirally-invariant ansatz for duality
with an infinite number of mesons. 

\vskip0.1in
\noindent {\twelvepoint{\it 7.2\quad A Minimal Realistic Model}}

\noindent One way to satisfy all constraints is to make an artificial separation between
the low-energy physics relevant to chiral symmetry and the high-energy physics
relevant to duality\ref{Srivastava}\ref{Golterman}. This requires introducing a discontinuity in
$n$. A simple ansatz\ref{Srivastava} is

$$\EQNalign{ 
&{{F_{\sss {V,A}}^2}}(n)= \cases{ {F_{{\rho},{a_1}}^2} & $n=0 $ \cr
                                   {{F_{\sss {V,A}}^2}} &  $n>0 $\cr}\; ,
\EQN simplean;a\cr
&{{M_{\sss {V,A}}^2}}(n)= \cases{ {M_{{\rho},{a_1}}^2} & $n=0 $ \cr
                                   {{M_{\sss {V,A}}^2}}+ {{\Lambda_{\sss {V,A}}^2}}(n-1) &  $n>0 $.\cr}
\EQN simplean;b\cr}
$$
Here we have extracted the lowest-lying vector and axial-vector mesons, $\rho$
and ${a_1}$, respectively. This is the minimal non-trivial model consistent
with chiral symmetry. The duality and chiral constraints then imply

$$\EQNalign{ 
&{{F_{\sss V}^2}}={{F_{\sss A}^2}}={\nc\over{24{\pi^2}}}{\Lambda^2}\; ,
\EQN sumcon;a\cr
&{{M_{\sss V}}}={{M_{\sss A}}}\equiv M \; ,
\EQN sumcon;b\cr
&{{\Lambda_{\sss V}}}={{\Lambda_{\sss A}}}\equiv\Lambda
\EQN sumcon;c\cr}
$$
where we have matched to the coefficient of the perturbative logarithm, and

$$\EQNalign{ 
&{F_{\rho}^2}-{F_{a_1}^2}=\yo1 \; ,
\EQN sumcon2;a\cr
&{F_{\rho}^2}{M_{\rho}^2}-{F_{a_1}^2}{M_{a_1}^2}=0.
\EQN sumcon2;b\cr}
$$ 
Notice that the vector and axial-vector mesons in the infinite
tower are degenerate.
With respect to the ${\Pi_{\sss{LR}}}({Q^2})$ correlator, this simple ansatz
has been investigated in many places\ref{sfsr}\ref{Peris}\ref{Beane2}. Here $\pi$,
$\rho$ and $a_1$, together with an isoscalar $S$,  fill out a reducible (10-dimensional)
$(\bf{1},\bf{3})\oplus(\bf{3},\bf{1})\oplus(\bf{2},\bf{2})$ representation,
while all other vector and axial-vector mesons are in irreducible
$(\bf{1},\bf{3})\oplus(\bf{3},\bf{1})$ representations.  It is interesting that
the chiral symmetry constraints effectively decouple the hadronic parameters
$M$ and $\Lambda$ from low-energy chiral physics$^6$\vfootnote6{\tenpoint The
authors of \Ref{Golterman} consider an ansatz given by \Eq{simplean} with
${{F_{a_1}^2}}=0$, match to the OPE and experience no such
decoupling. However, they do not impose the sum rules of \Eq{sfsr};
according to \Eq{sumcon} and \Eq{sumcon2}, consistency of their ansatz with chiral symmetry
requires ${{F_{\sss V}^2}}={{F_{\sss A}^2}}$, ${{M_{\sss V}^2}(n)}={{M_{\sss
A}^2}(n)}$ {\it and} ${M_{\rho}^2}=0$. In this case, $\pi$ and $\rho$ are in an
irreducible $(\bf{1},\bf{3})\oplus(\bf{3},\bf{1})$ representation.}. 
Inserting the ansatz, \Eq{simplean}, in \Eq{vectorsum}, 
doing the sums over $n$ and
matching to the OPE gives

$$\EQNalign{ 
&{\vev{{\cal O}}^{\sss d=2}_{\sss {V,A}}}=0=
2{F_{\rho}^2}-{\nc\over{12{\pi^2}}}({M^2}-{1\over 2}{\Lambda^2})\; ,
\EQN opejunk;a\cr
&{\vev{{\cal O}}^{\sss d=4}_{\sss {V,A}}}={{\alpha_s}\over{12\pi}}\vev{G_{\mu\nu}G_{\mu\nu}}=
-2{F_{\rho}^2}{M_{\rho}^2}+{\nc\over{24{\pi^2}}}({M^4}-{M^2}{\Lambda^2}+{1\over 6}{\Lambda^4})\; ,
\EQN opejunk;b\cr
&{\vev{{\cal O}}^{\sss d=6}_{\sss {V}}}=
-{28\over 9}{\pi{\alpha_s}}\vev{{\bar q} q}^2=
2{F_{\rho}^2}{M_{\rho}^4}-{\nc\over{36{\pi^2}}}
({M^2}-{1\over 2}{\Lambda^2})({M^2}-{\Lambda^2}){M^2}\; ,
\EQN opejunk;c\cr
&{\vev{{\cal O}}^{\sss d=6}_{\sss {A}}}=
{44\over 9}{\pi{\alpha_s}}\vev{{\bar q} q}^2=
2{F_{a_1}^2}{M_{a_1}^4}-{\nc\over{36{\pi^2}}}
({M^2}-{1\over 2}{\Lambda^2})({M^2}-{\Lambda^2}){M^2}
\EQN opejunk;d\cr}
$$ for the first few Wilson coefficients. One can develop a phenomenology for
the QCD condensates with this (or other) simple parametrizations of
duality. This is hampered by large uncertainties in the values
of the condensates. 
The relations of \Eq{sumcon2} can be parametrized by a single mixing angle, $\phi$, via
${F_\pi}={F_\rho}\sin\phi$, ${F_{a_{1}}}={F_\pi}\cot\phi$ and
${M_\rho}={M_{a_{1}}}\cos\phi$. 
The known vector excited states are $\rho' (1450)$,
$\rho'' (1700)$ and $\rho''' (2150)$\ref{pdg}.  Fitting to $\rho' (1450)$
we have $M=1450~{\rm MeV}$. Using \Eq{opejunk} with $F_\pi$, $M_\rho$ and $M$ as
input we then find $\Lambda =1189~{\rm MeV}$,
which predicts $M_{\rho''}=1875~{\rm MeV}$ and $M_{\rho'''}=2220~{\rm MeV}$.
These values differ from the experimental
values by amounts consistent with ${\cal O}(1/\nc )$ corrections. 
We also predict $\phi =44.4^{0}$, compared to
the value $\phi =37.4^{0}$ resulting from
fitting ${F_\rho}$ directly to ${\rho^{\sss 0}}\rightarrow{e^+}{e^-}$\ref{pdg}.
One then predicts ${F_{a_1}}=95~{\rm MeV}$ and ${M_{a_1}}=1078~{\rm MeV}$,
compared with the experimental values ${F_{a_1}}=122\pm 23~{\rm MeV}$ and
${M_{a_1}}=1230\pm 40~{\rm MeV}$. 
The predicted condensates are
${{\alpha_s}}\vev{G_{\mu\nu}G_{\mu\nu}}=0.06~{\rm GeV}^{4}$ and
${\pi{\alpha_s}}\vev{{\bar q} q}^2= 1.5\times 10^{-3}~{\rm GeV}^{6}$,
respectively. These values are somewhat large; recent determinations give  
${{\alpha_s}}\vev{G_{\mu\nu}G_{\mu\nu}}=0.048\pm 0.03~{\rm GeV}^{4}$~\ref{Indu}
and ${\pi{\alpha_s}}\vev{{\bar q} q}^2= 9\pm 2\times 10^{-4}~{\rm GeV}^{6}$~\ref{Stern}.

This model predicts excited axial-vector states with masses $M_{a1'}=1450~{\rm
MeV}$, $M_{a1''}=1875~{\rm MeV}$ and $M_{a1'''}=2220~{\rm MeV}$.  The particle
data group lists one excited axial-vector state, ${a_1}'(1640)$~\ref{pdg}.  The
splitting between this state and $\rho'(1450)$ is
consistent with an ${\cal O}(1/\nc )$ correction.  It will be very interesting
to have new data on the spectrum of excited vector and axial-vector mesons. It
is expected that the masses and widths of the low-lying excited vectors and
axial-vectors will be determined in the Hall D program at Jefferson Laboratory
in the near future\ref{Meyer}.

\vskip0.1in
\noindent {\twelvepoint{\bf 8.\quad Conclusion}}
\vskip0.1in

\noindent Two-point functions of conserved vector and axial-vector QCD currents
offer an interesting system to investigate quark-hadron duality. In the
large-$\nc$ limit, the duality matching conditions are tractable and, in
contrast with QCD in 1+1-dimensions, there are chiral symmetry constraints,
which take a particularly simple form.  Finding a {\it smooth} ansatz for
duality, consistent with all constraints, is equivalent to finding the infinite
dimensional matrix which mixes the irreducible chiral representations filled
out by the vector and axial-vector mesons. We find no smooth solution
consistent with the duality and chiral symmetry constraints when the vector and
axial-vector squared-masses are linear in $n$ and non-degenerate.  To avoid
this degeneracy it would appear necessary to go beyond the Regge-type
linear-spacing ansatz for the squared masses. In the large-$\nc$ limit, the
basic constraints of duality and chiral symmetry require vector-axial-vector
degeneracy in the meson spectrum at sufficiently high excitation
energy$^7$\vfootnote7{\tenpoint \Ref{Glozman} has come to similar conclusions
for not completely dissimilar reasons in the excited-baryon sector.}.  The
characteristic energy at which degeneracy should set in is
unknown. A simple realistic model, which predicts a tower of degenerate vector
and axial-vector mesons, is roughly consistent with existing data.

Although hadronic string models provide important insight into how correlators
determined by sums of infinite numbers of simple poles can be consistent with
chiral symmetry, they do not provide an easy analog which satisfies the
constraints of duality as well. Fundamentally this is because string models
exhibit Regge asymptotic behavior for four-point functions, which is governed
by fractional powers of the momentum transfer variable $Q^2$, while duality for
two-point functions involves the OPE, which does not see fractional powers of
$Q^2$. Hadronic string models do suggest simple generalizations which give
smooth solutions to the joint duality and chiral constraints in the degenerate
limit. However, the relation, if any, between these models and large-$\nc$ QCD
remains unclear.


\vskip0.1in
\noindent {\twelvepoint{\bf Acknowledgements}}
\vskip0.1in

\noindent 
I thank Martin Savage and Michael Strickland
for useful conversations, and Curtis Meyer for a valuable correspondence. 
This work was supported by the U.S. Department of Energy grant
DE-FG03-97ER-41014.

\nosechead{References}
\ListReferences \vfill\supereject \end